\documentclass{aa}  

\usepackage{graphicx}
\usepackage{txfonts}
\begin{document}

   \title{Active galactic nuclei and gravitational redshifts}


    \author{N. Padilla \inst{1}, S.~Carneiro \inst{2,3}, J.~Chaves-Montero \inst{4},  C.~J.~Donzelli \inst{1,5}, C.~Pigozzo \inst{2}, P.~Colazo \inst{1}, \and J.~S.~Alcaniz \inst{3}}
    \authorrunning{N. Padilla et al.}

   \institute{Instituto de Astronomía Teórica y Experimental (IATE), CONICET - U. Nacional de Córdoba, X5000BGR, Córdoba, Argentina\\
              \email{nelson.padilla@unc.edu.ar}\\
          \and
             Instituto de Física, Universidade Federal da Bahia, 40210-340, Salvador, BA, Brazil\\
          \and
            Observat\'orio Nacional, 20921-400, Rio de Janeiro, RJ, Brazil\\
          \and
             Institut de F\'isica d’Altes Energies, The Barcelona Institute of Science and Technology, Campus UAB, E-08193 Bellaterra (Barcelona), Spain\\
          \and
            Observatorio Astron\'omico de C\'ordoba, Universidad Nacional de C\'ordoba, X5000BGR, C\'ordoba, Argentina
             }

   \date{Received September 26, 2023; accepted ...}

 
  \abstract
   {Gravitational redshift is a classical effect of Einstein's General Relativity, already measured in stars, quasars and clusters of galaxies.}
   {We here aim to identify the signature of gravitational redshift in the emission lines of active galaxies due to supermassive black holes, and compare to what is found for inactive galaxies.}
   {Using the virial theorem, we estimate gravitational redshifts for quasars from the 14th data release of the Sloan Digital Sky Survey, and compare these with measured ones from the difference between the redshifts of emission lines of Sydney Australian Astronomical Observatory Multi-object Integral Field (SAMI) galaxies in central and outer annuli of their integral field spectra.}
   {Firstly, from the full width at half maximum of 
   {$H_\beta$} lines of 57 Seyfert type I galaxies of the AGN Black Hole Mass Database, we derive a {median gravitational redshift $z_g = 1.18 \times 10^{-4}$. Expanding this analysis to 86755 quasars from DR14 of SDSS we have a median value $z_g =  1.52 \times 10^{-4}$}. Then, by comparing the redshifts of 
   {$34$} lines measured at  central and outer regions of LINER galaxies in the SAMI survey we obtain $z_g = (0.68 \pm 0.09) \times 10^{-4}$, {which increases to $z_g = (1.0 \pm 0.1) \times 10^{-4}$ when using $H_\alpha$ and $H_\beta$ lines}. These numbers are compatible with central black holes of $\approx 10^9$ solar masses and broad line regions of $\approx 1$~pc. For non-AGN galaxies the gravitational redshift is compatible with zero.}
   {}

   \keywords{Supermassive black holes -- gravitational redshift -- emission lines -- cosmological parameters
               }

   \maketitle
%

\section{Introduction}

 Gravitational redshifts have been already measured in stars, quasars \citep{Mediavilla_2018,Mediavilla} and galaxy clusters \citep{Wojtak, Mpetha, Rosselli}. 
A white dwarf of Chandrasekhar mass, for example, has a radius bounded by $R < 0.01\, R_\odot$, which leads to a lower limit for its gravitational redshift $z_g \approx 1.4 \times 10^{-4}$. 
In the case of galaxies, if they are inhabited by supermassive black holes (SMBH), their gravitational redshift can affect light emitted from the galactic bulges, and, as we will see below, the gravitational redshift can also be as high as $10^{-4}$ in the case of active galactic nuclei (AGN) hosts.

The signature of gravitational redshifts in galaxy emission lines would constitute an additional confirmation of General Relativity, and it was already detected, for instance, in \ion{Fe}{iii}$\lambda\lambda$2039–2113 lines of quasars \citep{Mediavilla_2018,Mediavilla}. {Gravitational redshift in quasar emission lines has also been reported in \cite{Bon}, and its signature was also find in iron K$\alpha$ \citep{Nandra} and H$\alpha$ \citep{Da-in} emission lines of Seyfert type 1 galaxies. }

The present work is also an attempt to show a signature of SMBHs gravitational redshift in galaxy {emission lines, 
concentrating mostly on LINER and non-AGN galaxies, by directly comparing the measured redshifts of central and outer regions of extended galaxies.  Taking into account possible sources of systematics we  infer gravitational redshifts for these galaxies and we  compare them with those estimated for Quasars and Seyfert type I galaxies}. 

{This work is organised as follows.  In Section \ref{sec:2}}, we first compute, indirectly, the gravitational redshift of Seyfert type I galaxies of the AGN Black Hole Mass Database \citep{Bentz}, from the full width at half maximum (FWHM) of hydrogen lines. The study is then extended to about eighty thousand quasars of the 14th data release of the Sloan Digital Sky Survey. In Section 3 a direct measurement is performed through a comparison of redshifts of lines emitted from the outer and central regions of galaxies of the Sydney Australian Astronomical Observatory Multi-object Integral Field \citep[SAMI;][]{Croom} catalogue, {as we expect the effect of the gravitational redshift of a central SMBH to contaminate only the central regions of well resolved galaxies. }{ Section 4 studies whether the systematic error introduced by the gravitational redshift of the SMBH in AGN galaxies can have an impact on cosmological parameter estimations}.  In the last section we discuss the results and outline our conclusions.

\section{Gravitational redshifts of AGN and quasars}\label{sec:2}

We begin with a statistical analysis of the gravitational redshifts of active galactic nuclei (AGNs) inferred from the full width at half maximum of the 
$H_{\beta}$ emission lines. Applying Kepler’s law to the accretion disc of a central SMBH of mass $M$, one finds the following gravitational redshift {\citep{Bentz}},
\begin{equation} \label{1}
z_g\equiv\frac{GM}{Rc^2} = f \left( \frac{v}{c} \right)^2,
\end{equation}
where $R$ is the distance from the emission region to the central SMBH and $v/c$ is the rotational linear velocity of gas in that region in units of the speed of light, which is typically measured from the FWHM of the emission line. The statistical factor $f$, typically of order unity, is related to uncertainties in the distribution and shape of the emitting clouds and relativistic corrections.

\begin{figure}
    \centering
    \vskip -.1cm
    \includegraphics[scale=.4]{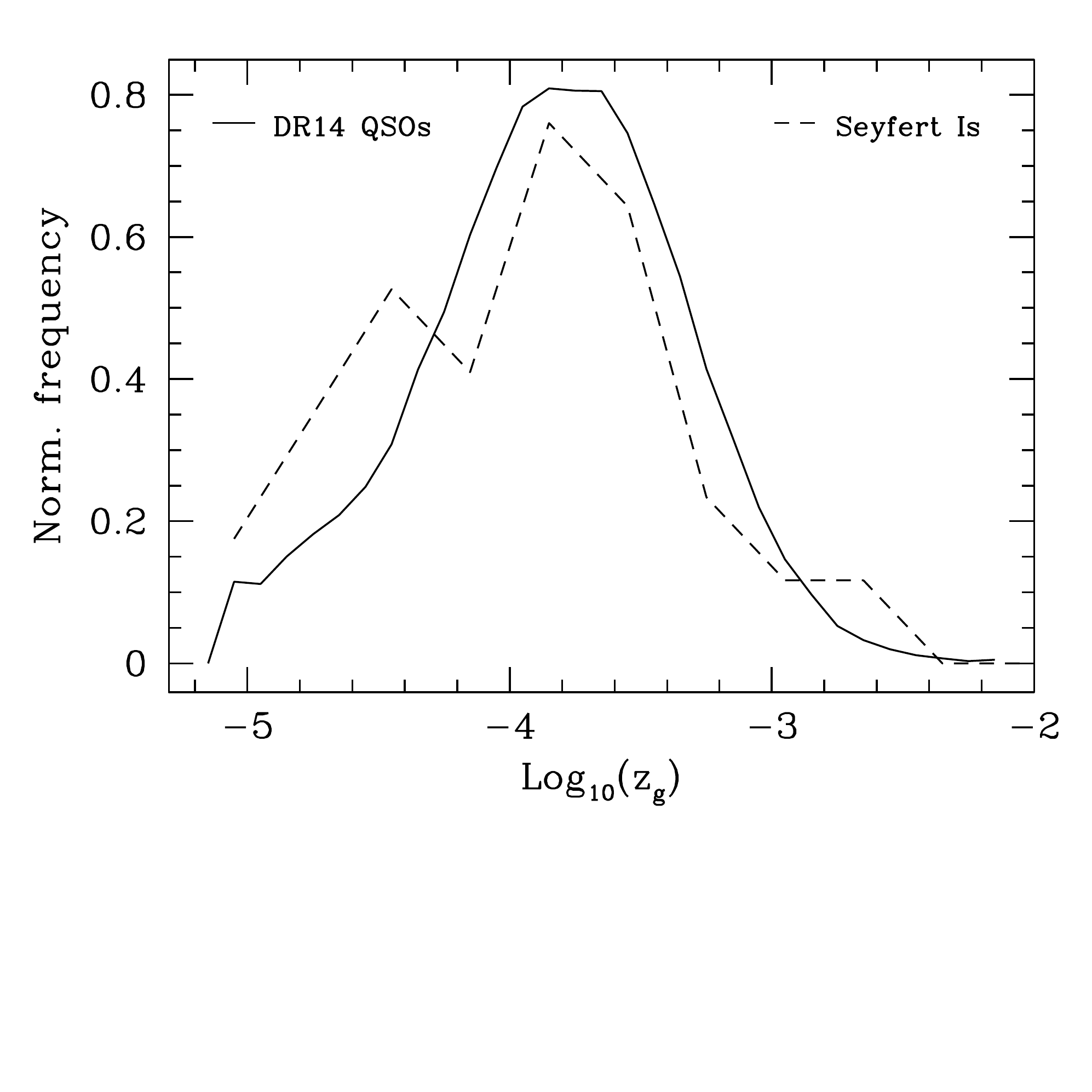}
    \vskip -2.0cm
    \caption{{Normalised distributions of gravitational redshifts indirectly derived from the full width at half maximum (FWHM) of $H_{\beta}$ broad lines of SDSS DR14 quasars \citep{Stalin} (solid), and of  {$H_{\beta}$} broad lines of $57$ Seyfert type I galaxies catalogued in The AGN Black Hole Mass Database \citep{Bentz} (dashed lines).  The median and dispersion of the distribution for the SDSS DR14 quasars are median$(z_g)=1.52\times 10^{-4}$,  and Std$(\log_{10}(z_g))=0.49$, whereas for the AGN Black Hole Mass Database, {median$(z_g)=1.18\times 10^{-4}$,  and Std$(\log_{10}(z_g))=0.58$}.}  
    } \label{histogram} 
   \end{figure}

{In this way, by directly computing the right-hand side of Eq.~(\ref{1}) from the measured FWHM,} we show in Fig.~\ref{histogram} the distribution of gravitational redshifts for {57 Seyfert I galaxies inferred from the FWHM measurements of $H_{\beta}$ lines of the AGN Black Hole Mass Database \citep{Bentz} for $f = 1$ (dashed line), for which we find a median value of $z_g \simeq 1.18 \times 10^{-4}$.} 
The solid line shows the analysis of 86755 quasars from DR14 of SDSS with FWHM measurements of $H_{\beta}$ \citep{Stalin}, finding a median value $z_g \simeq  1.52 \times 10^{-4}$.  In both cases the dispersion around the typical inferred gravitational redshifts is of about the same factor, 
reflecting a similar variety of AGNs in both samples. 
These figures should still be considered with caution because of the uncertainties involved in {the 
use of the FWHM of hydrogen lines to estimate central black hole masses \citep[e.g.,][]{Marziani}.}

{The broad line region dominates the luminosity in the centre of broad line AGNs and QSOs, where spectroscopy fibres are usually placed to measure galaxy redshifts.  This suggests that broad line AGN and QSO redshifts are biased due to the SMBH gravitational redshift.}

{For galaxies with lower or absent nuclear activity (narrow line AGN, LINERs, non-AGN galaxies), this contamination can be even negligible, especially if the galaxy has no active nucleus which represents the majority of the cases.  But it could still be significant in narrow-line AGN/LINERs. 
Here we present an attempt to directly identify a  signature of gravitational redshift in the latter types of galaxies, which} is, {\it per se}, of astrophysical interest.

\section{Gravitational redshift in emission lines}

With this goal in mind, we use a compilation of galaxies from the third data release of the Sydney Australian Astronomical Observatory Multi-object Integral Field \citep[SAMI;][]{Croom} to directly measure the effect of the gravitational redshift of the region surrounding SMBHs, 
particularly when looking central and outer regions of galaxies.  

{Since the SAMI galaxy survey is an Integral Field Spectroscopy survey, the galaxies in our sample are nearby and well resolved spatially}, which allows us to obtain high signal-to-noise spectra of extended galaxies.  {The SAMI spectrograph has two arms for blue and red wavelengths, spanning the ranges $3700-5700$ and $6250-7350$ Angstr\"oms, with a resolution of $\sigma=70$ km/s and $\sigma=30$ km/s, respectively. } SAMI provides datacubes of 15 arcsec diameter in an array of $50$x$50$ interpolated spaxels (spectra per projected pixel), which are further processed into annularly averaged spectra at different radii from the galaxy centre by the SAMI team, with a central spectrum averaged within a disc of 1.5 arcsec radius (equivalent to $1.3$~kpc at the median redshift of the sample $z=0.043$), and within annuli increasing in radius by 3 arcsec intervals.  

We measure the redshift for individual emission lines in the central annulus and in the third annulus centred at 7.5 arcsec, far enough from the centre to avoid contamination from it by {atmospheric} seeing and the point spread function of the observations, but with enough signal to noise ratio to be able to confidently detect emission lines. For simplicity, in what follows we will refer to the third annulus as the outer one.
\begin{figure}
    \centering
    \vskip -1.cm    \includegraphics[width=9cm]{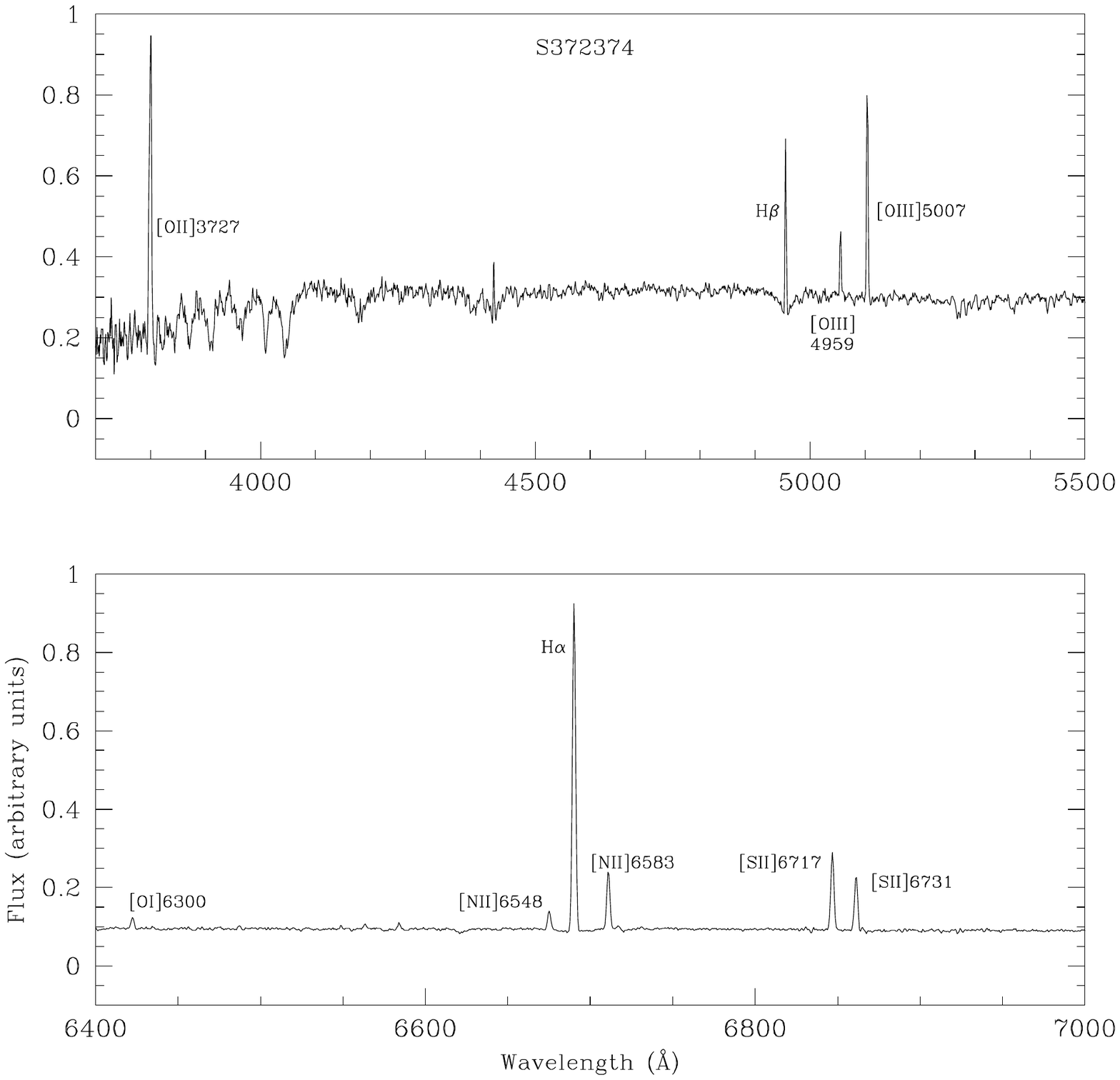}
    \vskip -3cm    \includegraphics[width=9cm]{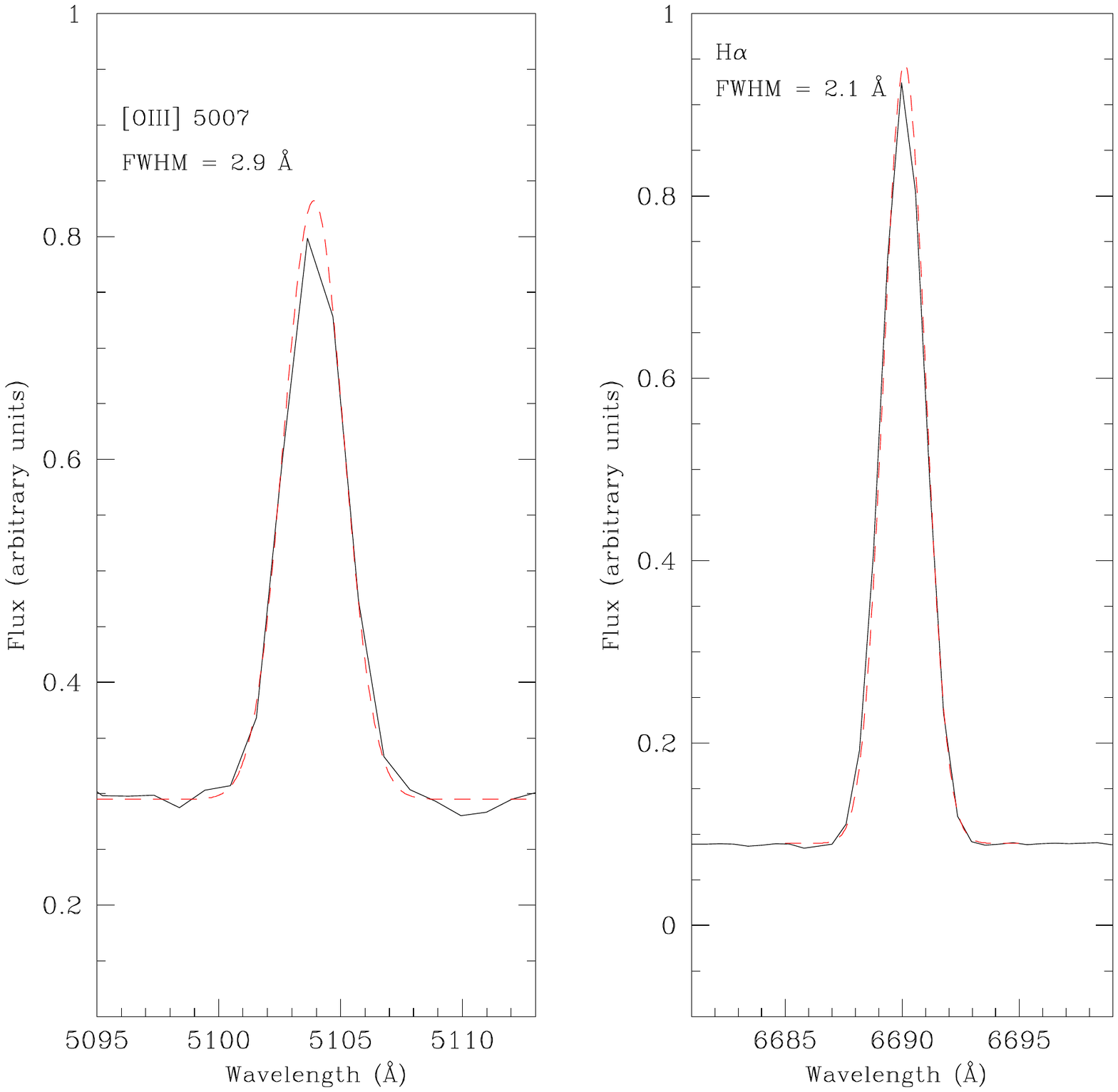}
    \vskip -2.3cm
    \caption{ Example spectrum for galaxy S372374 in the blue (top) and red arms of the SAMI spectrograph (middle). The different lines used in our analysis are indicated in the spectrum.  The bottom panel shows two example fits {(red dashed lines)} done using Gaussian functions {via the SPLOT routine in IRAF}.}\label{fig:spectra}
   \end{figure}

We concentrate on galaxies with large axial ratios (face-on galaxies with $\left< b/a \right> \simeq 0.8$) since we are interested in annularly averaged spectra.  {The spectra of galaxies selected this way} are more reliable and have higher total line luminosities. Whenever possible, we also restrict our selection to galaxies with detectable line emission out to the outer limits of the SAMI field of view to be able to measure redshifts on the external parts of galaxies {(up to the fourth annulus)}. Since AGN galaxies are less frequent, we relax these criteria to allow a larger number of AGN galaxies. We additionally require that the emission lines of all galaxies that enter our SAMI sample are well fitted by single Gaussians, i.e., with as little systematic effects from galaxy kinematics as possible, as the latter frequently produce composite or double emission lines which can be especially prominent further away from the galaxy centre and affect FWHM measurements. Our final sample contains 40 galaxies, of which  
9 show clear AGN signatures according to the \citet{BPT} diagram using the \citet{kauffmann03} limits (out of which 8 are LINERs and 1 is a Seyfert II galaxy), 9 are starbursts (HII) and the rest are composite galaxies{, i.e. starbursts + AGN, although we will refer to these simply as non-AGN}.
Given the multiple and flexible selection criteria we needed to adopt to construct these samples, we provide the list of SAMI IDs and the classification diagrams
in the Appendix. {We employed the SPLOT routine within IRAF to fit} a total of 306 emission lines that can be detected both in the central and outer annuli of our sample.

We fit Gaussians to the individual lines [OII]$3727$, $H_{\beta}$, [OIII]$4959$, [OIII]$5007$, [OI]$6300$, [NII]$6548$, $H_{\alpha}$, [NII$6583]$, [SII]$6717$, and [SII]$6731$, whenever any of these is available in a spectrum. Fig. \ref{fig:spectra} shows an example spectrum for galaxy S$372374$ in the two top panels, for the blue and red arms of the spectrograph, respectively.  The bottom panel shows two examples of fits to  [OIII] and $H_{\alpha}$ lines, with the recovered values of FHWM.

The $z_g$ is found by subtracting the redshift measured at the {outer annulus of these AGN and non-AGN galaxies samples from those measured in the central annulus, using the same} rest-frame wavelength emission lines. 

We adopt this method 
since it allows us to estimate errors statistically as shown below.  We also check that using the average inner and outer redshifts over all available lines provides consistent results.  Since the $z_g$ for the Seyfert II galaxy {in our sample} is consistent with zero, as expected for this type of AGN galaxy, we will only consider LINERS as our AGN sample.

\begin{figure}
    \centering
    \vskip -0.3cm    \includegraphics[width=\columnwidth]{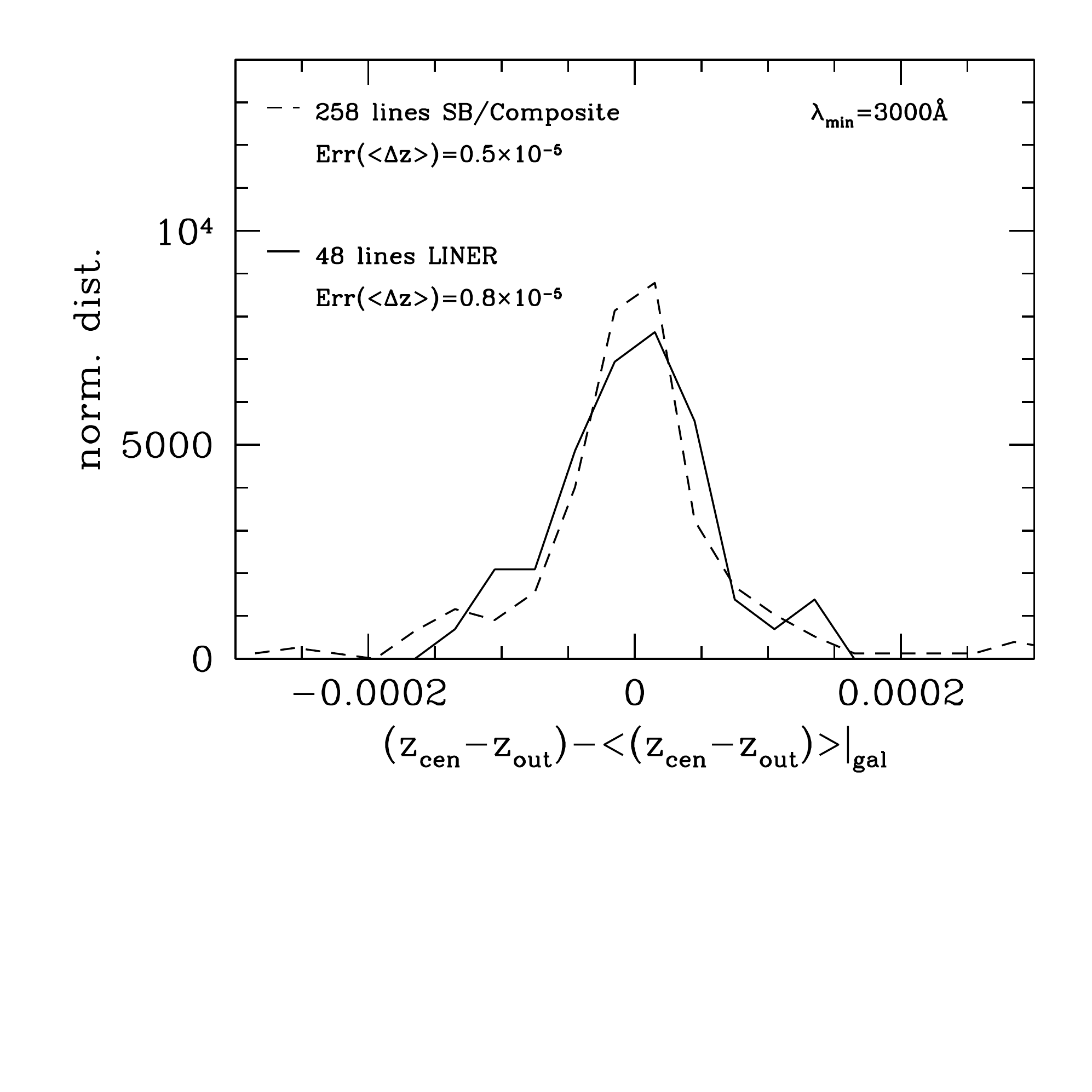}
    \vskip -2.7cm    \includegraphics[width=\columnwidth]{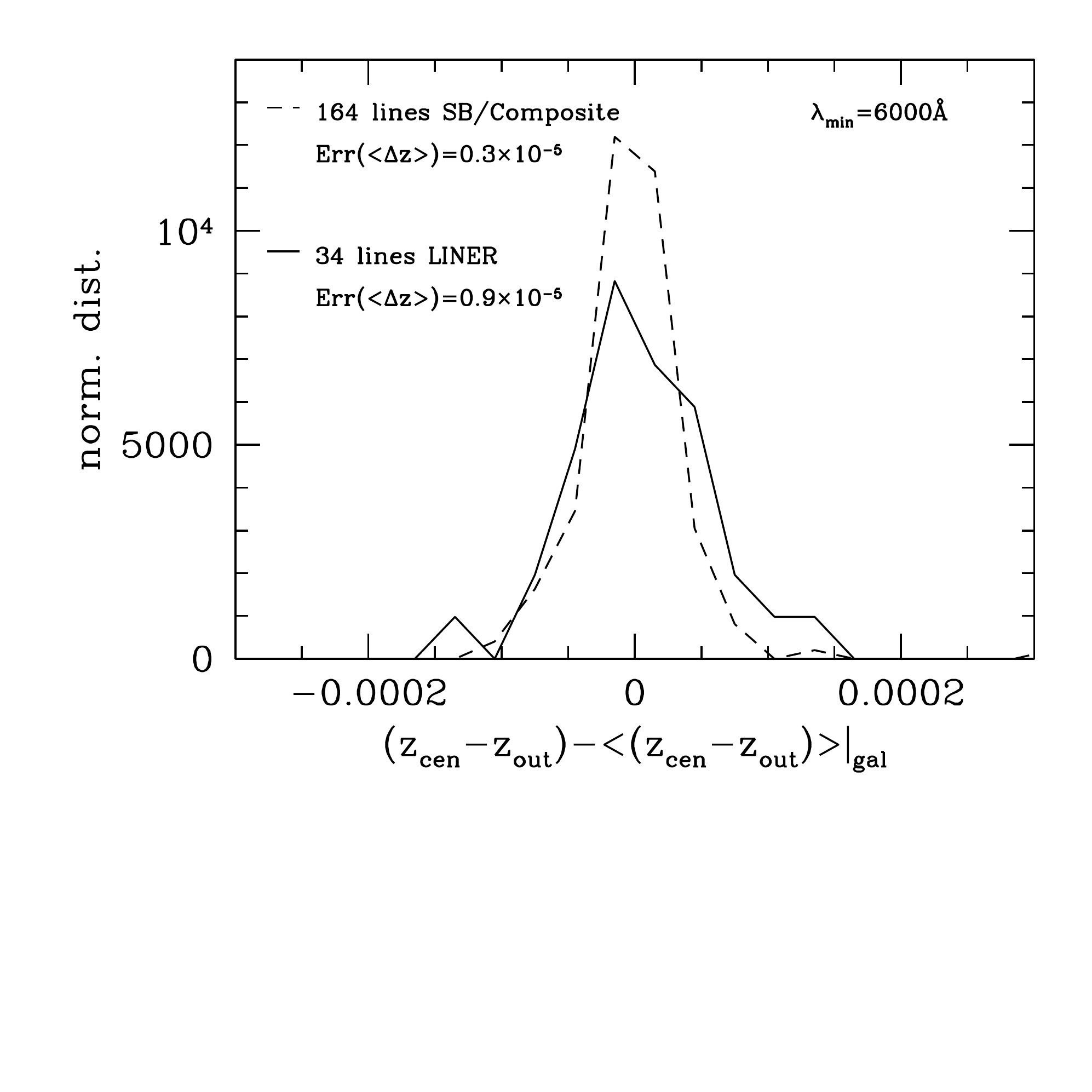}
    \vskip -2.7cm
    \caption{{\bf Top:} Distribution of variation of $z_g$ for AGN {(LINER)} and non-AGN {(SB/composite)} lines within individual galaxies. {The legend shows the error on the mean of the distribution, which we assign as the error in the measured redshift differences for the different samples.} {\bf Bottom:} The same when only red lines, {with better spectral resolution,} are included.}\label{histogram3}
   \end{figure}
   
\begin{figure}
    \centering
    \vskip -0.3cm    \includegraphics[width=\columnwidth]{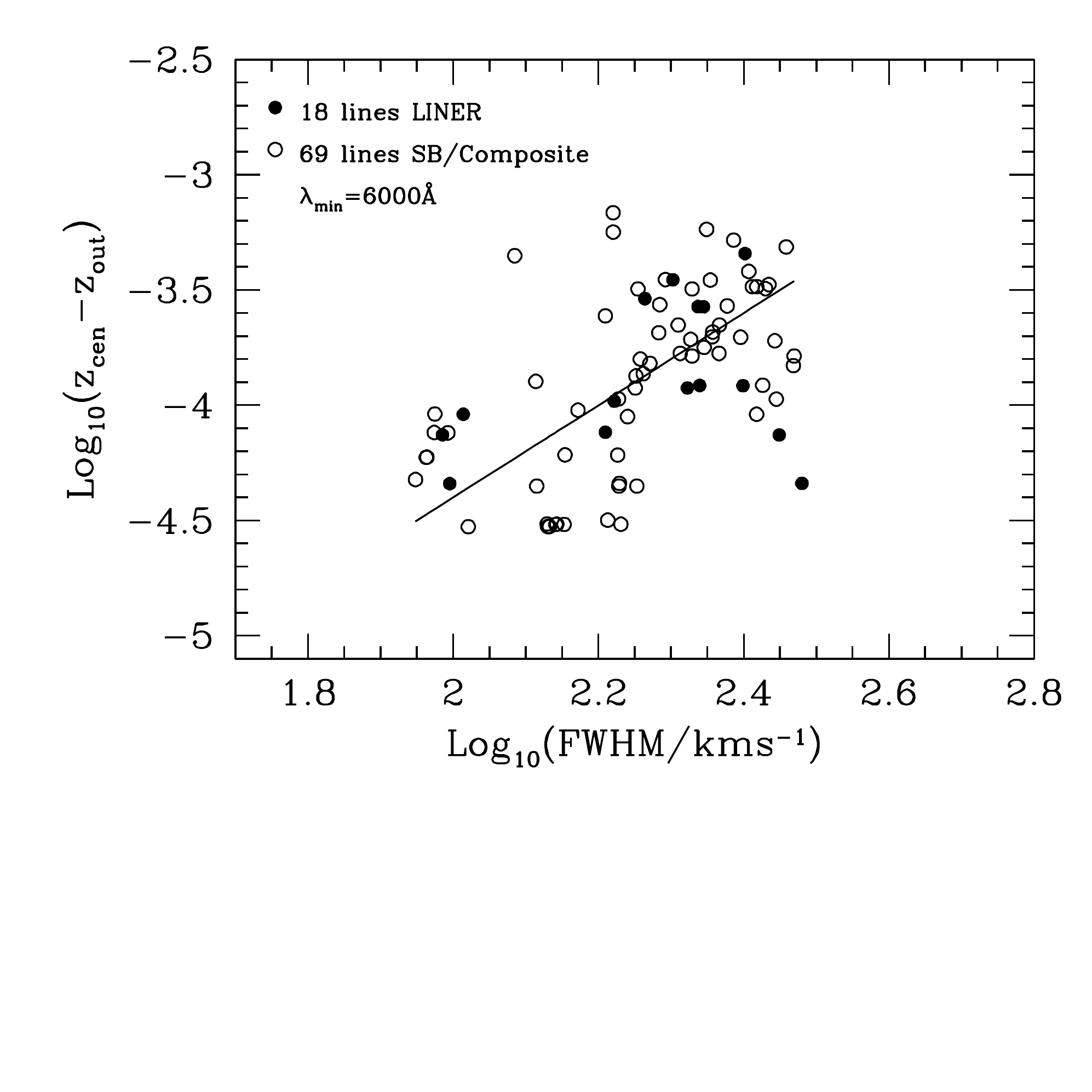}
    \vskip -2.7cm
    \caption{Correlation $z_g \times \text{FWHM}$ for {LINER} (solid dots) and {SB/composite} (empty dots) galaxies. The gravitational redshift $z_g$ is given by the difference $z_{\text{central}} - z_{\text{out}}$ between the redshifts of each line measured at the central and outer regions. FWHM's are measured at the {inner annulus}. The line shows a fit following Eq.~(1) { where only the amplitude is fitted to the points}, {corresponding to $f_{\rm fit}\sim 400$}.}\label{correlation}
   \end{figure}

\begin{figure}
    \centering
    \vskip -0.3cm
    \includegraphics[scale=.44]{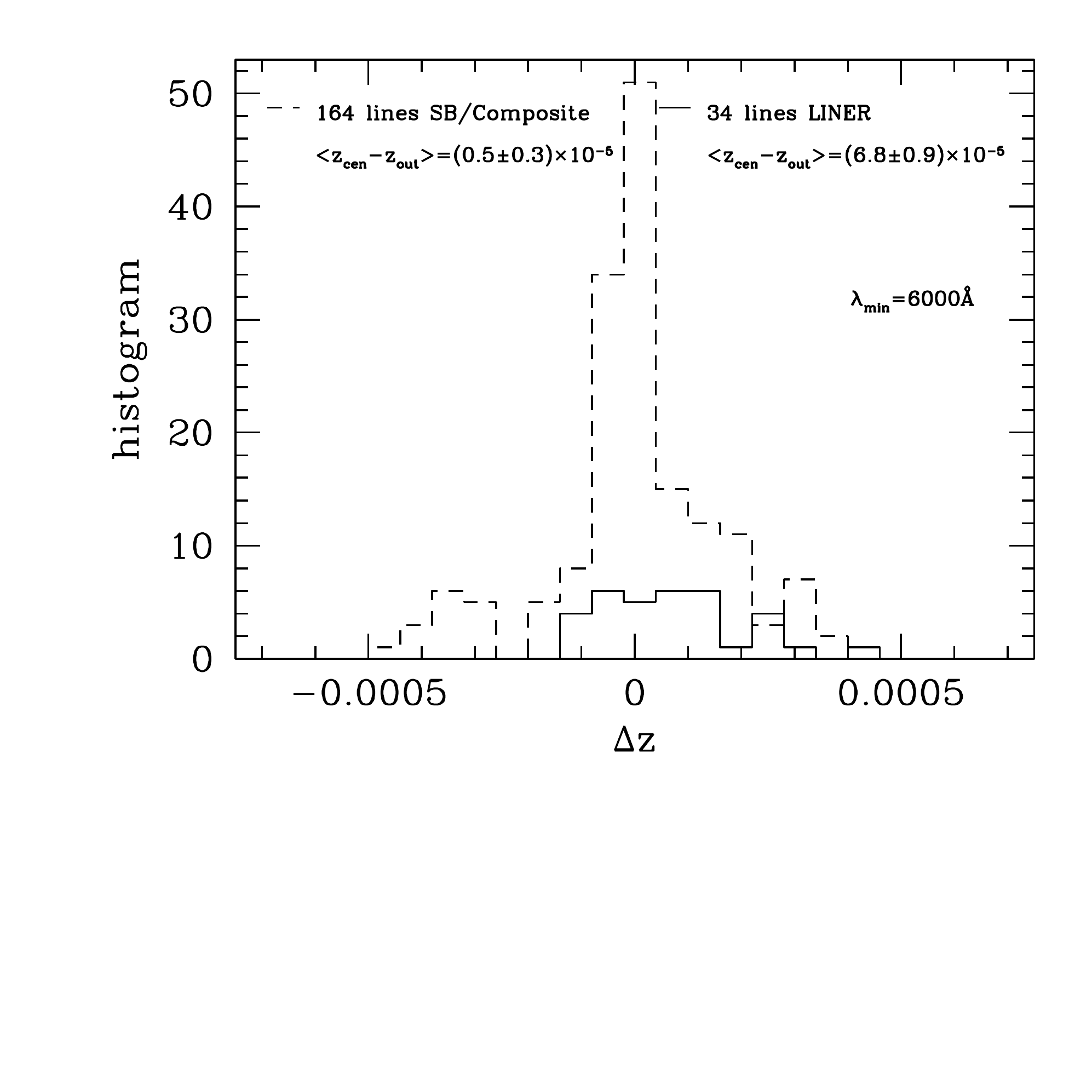}
    \vskip -2.7cm
    \includegraphics[scale=.44]{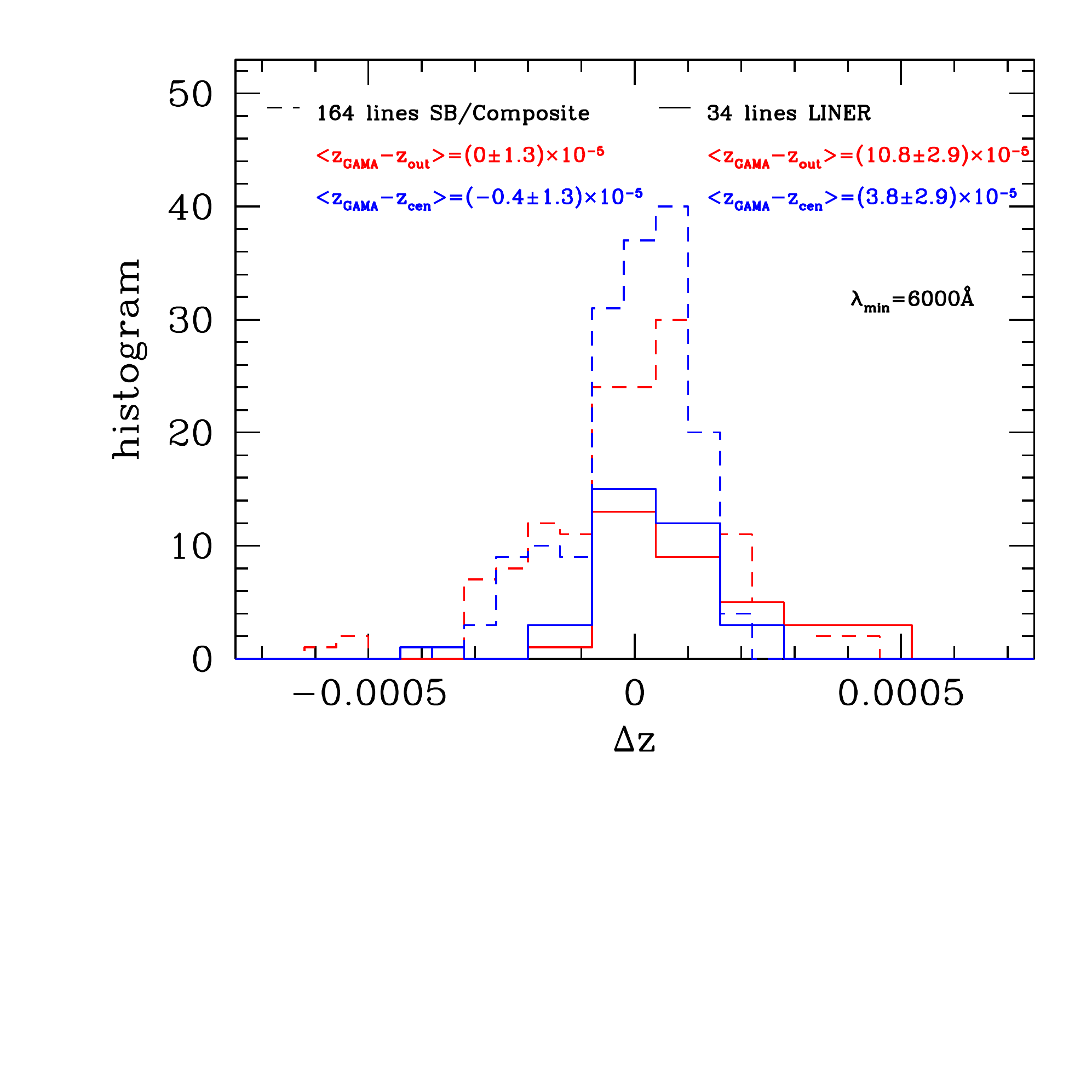}
    \vskip -2.7cm
    \caption{Histograms of $z_{\text{central}} - z_{\text{out}}$ for {LINER} and non-AGN galaxies (upper panel). In the lower panel we also show the histograms for $z_{\text{GAMA}} - z_{\text{out}}$ and $z_{\text{GAMA}} - z_{\text{central}}$.  {In this panel the size of the bin for the GAMA LINER comparison is twice as large than the other histograms in this figure to reduce noise.}}\label{histogram4}
   \end{figure}

We {devise a method to estimate errors that takes advantage of multiple measurements of the redshift of the same emission line in individual galaxies.  We also compare the resulting galaxy central redshifts with those quoted by the GAMA survey as a consistency check. We } estimate the {error of our measured gravitational redshifts $\Delta z=z_{\text{central}}-z_\text{{outer}}$ by calculating this difference for different emission lines and measuring the standard deviation within individual galaxies.   We do this} only when single galaxies allow measurement of { multiple emission lines of equal rest frame wavelength in both the inner and outer annuli.  The histograms of Fig.~\ref{histogram3}} show the distribution of deviations of $\Delta z$ {around the mean obtained from emission lines in individual galaxies}  in our AGN and non-AGN samples (solid and dashed lines, respectively).  {The legend shows the estimated error calculated by dividing the dispersion of $\Delta z$ by the square root of the number of galaxies where this measurement was possible.} As the resolution of the red arm of the SAMI spectroscopic instrument is {about} twice that of the blue arm, it is not surprising that when only red lines are considered (lower panel), { the error of $\left<\Delta z\right>$} is lower than that derived with blue and red lines together, {at least for the more numerous non-AGN galaxy sample}. For this reason we have included only lines with wavelength above $6000$ Angstr\"oms in our analysis. 

We show in Fig.~\ref{correlation} the relation between $\Delta z$ and FWHM of emission lines measured in the inner annulus (solid dots indicate AGN galaxies, open circles, non-AGN galaxies).  
Note that {broader lines show a clear tendency to have a larger redshift difference which, for the reasons outlined below, we interpret as gravitational redshift contamination in the nucleus of order $10^{-4}$.  This tendency is only slightly sharper for AGN galaxies {(The dispersion around the fit for LINERs is about half that for SB for FWHM$<250$km$/$s)}. These data } can be reasonably well fitted by a straight line with slope $2$, i.e., the theoretical dependence $z_g \propto v^2$ given in Eq.~(\ref{1}). {We are unable to perform a fully automatic fit on these data given the large dispersion in the distribution.  However, we do fit the amplitude which we will refer to in our discussion below.}

Fig.~\ref{correlation} {is restricted to cases where a positive gravitational redshift signature is found but as we show next there is an important fraction of cases where} the redshift is higher in the outer region. 

{We now focus on the measured gravitational redshifts in our SAMI samples.  The upper panel of Fig.~\ref{histogram4} shows $\Delta z=z_{\text{central}}-z_\text{{outer}}$ for AGN and non-AGN galaxies.  Interpreting this redshift difference as {being exclusively of gravitational origin, the mean gravitational redshift of the AGN sample (solid line) is} 
$z_g = (0.68 \pm 0.09) \times 10^{-4}$. 

On the other hand, for non-AGN galaxies (dashed line) we obtain a result almost compatible with zero}. In fact, no significant gravitational redshift is expected for the latter. 
{Indeed, 
in the absence of an AGN, the effect of the galaxy's gravitational redshift is extended over the entire galaxy. As the ratio between the BLR radius and the half-light radius is typically $< 10^{-3}$, while the ratio between the central SMBH mass and the mass of the host bulge is typically $\sim 10^{-2}$ \citep{1998AJ....115.2285M}, the resulting redshift is, at least, an order of magnitude lower. That is, the gravitational redshift for a typical non-AGN galaxy 
is $< 10^{-5}$, below the precision we can achieve, and this number is not altered even if we include a non-active SMBH.} 

{Our interpretation of $\Delta z$ as a gravitational redshift is further supported by the  redshift differences between the third and fourth annuli which we measure whenever possible.  In this case we find}  that the mean values are compatible with zero for both AGN and non-AGN. This result confirms that the redshift difference has its origin in the central region. 
The use of the second annulus, on the other hand, has not led to conclusive {results because of its complex kinematics, affected simultaneously }by both the nucleus and the galactic disc.

The parent galaxy sample of SAMI is the Galaxy and Mass Assembly catalogue \citep[GAMA;][]{Liske}, which allows us to compare our  {IRAF-SPLOT obtained} estimates of SAMI galaxy redshifts with the automated redshift measurement of the GAMA survey.  GAMA uses a single fibre spectrograph of $2$ arcsec diameter, slightly smaller than the central $3$ arcsecs disc used in our 
SAMI calculations. In the lower panel of Fig.~\ref{histogram4}, we also present histograms for the differences $z_{\text{GAMA}} - z_{\text{out}}$ and $z_{\text{GAMA}} - z_{\text{central}}$, where $z_{\text{GAMA}}$ is measured with a central single fibre. As can be seen, our SAMI central measurements {are consistent} with GAMA redshits, and 
the offset between central and outer redshifts we find in SAMI {is confirmed. These redshift difference measurements are summarised in Table \ref{table:1}.} {Note that, for non-AGN galaxies, the SAMI central redshifts and the GAMA redshifts are consistent within the errors, whereas for AGN galaxies SAMI's is a factor $\approx 2/3$ lower, and just outside the estimated error. Since the GAMA fibres cover a nuclear area $(2/3)^2$ smaller than the SAMI central annulus, this small difference in the measured gravitational redshift is in fact expected for AGN if there is a systematic effect from the SMBH $z_g$.}

These results suggest that the SMBH gravitational redshift is negligible for non-AGN galaxies  
which {correspond to the majority of our galaxies as expected, as these are more common than AGN/LINERs} \citep{Bongiorno}. On the other hand, for AGN galaxies we find a gravitational redshift signature consistent with the {analysis of FWHM of broad line AGNs and quasars of Section \ref{sec:2}}.
   
{We also check the consistency of our measurements by using the redshifts of} only $H_{\alpha}$ and $H_{\beta}$ lines.  In this case we obtain a slightly higher mean value of the systematic for AGN galaxies, $z_{\text{central}} - z_{\text{out}} = (1.0 \pm 0.1) \times 10^{-4}$. As these lines are mainly emitted from the broad line region, this result strongly suggests that we are indeed measuring a gravitational redshift and not an unknown systematic.  

\begin{table*}
\caption{{Redshift differences between central and outer annuli, as obtained from SAMI data, and from adopting the GAMA redshift as the central one.  The fourth column shows that the redshifts of the central SAMI annulus and of GAMA are similar.}}             
\label{table:1}      
\centering                          
\begin{tabular}{c c c c c }        
\hline\hline                 
Galaxy type & Central minus outer & GAMA minus outer &GAMA minus central & \# of lines  \\    
\hline                        
   SB/Composite & $(0.5\pm0.3)\times 10^{-5}$ & $(0\pm1.3)\times 10^{-5}$ & $(-0.4\pm1.3)\times 10^{-5}$ & 164 \\ 
   LINER & $(6.8\pm0.9)\times 10^{-5}$ & $(10.8\pm2.9)\times 10^{-5}$ & $(3.8\pm2.9)\times 10^{-5}$ & 34 \\ 

\hline                                   
\end{tabular}
\end{table*}

{\section{Cosmological consequences of gravitational redshifts due to SMBHs}}

{The estimation of cosmological parameters from supernovae analyses could be affected by different systematics \citep{2011PhRvD..84h3005D, davis15, davis17}. For example, \cite{davis17} fitted the Joint Light-curve Analysis compilation \citep[JLA;][]{JLA} considering a systematic error $z_g$ as a free parameter and found that $z_g \approx (2.6 \pm 2.8) \times 10^{-4}$ alleviates tensions surrounding the inferred matter density parameter $\Omega_m$. The impact of such a $z_g$ on the inferred Hubble parameter was investigated in turn by \citet{Carneiro}, who find that it leads to an enhanced uncertainty in $H_0$ that could alleviate the current observational tension between the local expansion rate \citep{riess19} and the value of $H_0$ derived from the CMB anisotropy spectrum assuming the standard  $\Lambda$-Cold Dark Matter ($\Lambda$CDM) model.}

{Our measurement of $z_g$ for LINERs, as well as those estimated from the FWHM of Seyfert Is and QSOs (cf. Figure \ref{histogram}) fall short by about a factor of $2$ of the one proposed by \cite{davis17} as a possible solution to cosmological tensions; additionally, the fraction of AGN in galaxies is known to be low.  However, it is still interesting to check whether this could have any impact  on the SNe analysis.  We estimate the fraction of AGN in the JLA SNe hosts, searching publicly available catalogues using the SIMBAD database\footnote{https://simbad.cds.unistra.fr/simbad/tap/tapsearch.html} by object name when available or by restricting the angular separation to $60$ kpc in projection and to a redshift within $10\%$ of that of the SNe (with a maximum redshift difference set to $0.04$), when the object name was not available.  The search produced a total of 131 multiple matches and 494 unique ones. The multiple matches were visually inspected and resulted in 103 secure matches. The single matches were also inspected visually in some cases. Matches corresponding to non-public JLA Sloan Digital Sky Survey data had to be discarded. This procedure resulted in a total of $583$ total matches with available AGN classification. Of these, only $5$ are Seyfert I (broad line AGN) and $28$ are other types of AGN including LINER and Seyfert II galaxies, i.e. roughly $\sim 5\%$ of SNe hosts are AGN.  This percentage is low enough that the SMBH gravitational redshift is not likely to affect cosmological inferences from SNe.}

{
 On the other hand, the use of quasars as standard candles has recently shown promise \citep{Risaliti:2018reu,Sacchi:2022ofz}, and it is argued that QSO Hubble diagrams can be precise enough to constrain, for instance, the matter density parameter \citep{Dainotti:2023cpn}. Although  quasar redshifts are generally higher than those of SNe Ia samples and the gravitational redshift found here is smaller than the systematic proposed by \cite{davis17} in JLA, its possible effect on the precision of the matter density parameter determination from QSO Hubble diagrams would deserve further attention.
}

\vline

\section{Concluding remarks}

Our aim was to use emission lines in the nuclei and outskirts of AGN and non-AGN galaxies to find signatures of gravitational redshifts produced by the central supermassive black holes in galaxies, { and to compare these to gravitational redshift estimates obtained from measured broad line widths.}  

{We measured redshifts from the inner and third annuli of SAMI galaxies corresponding to the central $1.5$ arcsec disc and to the annulus between $4.5$ and $7.5$ arcsec.  We chose well resolved face-on galaxies with emission lines with single Gaussian profiles, with and without AGN activity.} {Our measurements of central galaxy redshifts are consistent with those reported by the GAMA survey.  }

The highest average signature of gravitational redshift we found for narrow line AGN hosts is $z_{\text{GAMA}} - z_{\text{out}} \approx 1.1 \times 10^{-4}$, similar to what we find when using only $H_{\alpha}$ and $H_{\beta}$ lines. Using all AGN SAMI lines in the red arm, we obtained 
$z_g \approx 0.7 \times 10^{-4}$ from $z_{\text{central}} - z_{\text{out}}$ (see Fig.~\ref{histogram4}). 
These values are {comparable to those inferred   from the analysis of FWHM's of broad line AGNs and quasars performed in Section 2, and are} compatible with the gravitational redshift of a central SMBH of $10^8 - 10^{9}$ solar masses with a broad line region of radius $0.1 - 1$pc.


{Our measured gravitational redshifts are subject to possible contamination  by a 
starburst contribution (SB), 
since star} formation can mask away broad line region (BLR) lines. \citet{Pastoriza} showed that a 30\% contamination of starburst lines is enough to make an AGN appear as non-AGN in the diagnostic diagrams, i.e., {to mask out a} BLR even when one is present. In particular, when looking at non-AGN galaxies, one could be looking at combinations of starbursts plus a faint AGN. 
The contamination would be present in both cases, but it would be more significant in the non-AGN. Another possible {source of} contamination of the central annuli {could come from} blueshifted outflows \citep[e.g.,][]{outflow}, which could also contribute to the observation of lines with negative values of $z_{\text{central}} - z_{\text{out}}$ {in addition to measurement uncertainties of the line redshifts}.



{The galaxies in our SAMI sample are 
not extreme AGNs; we have only $8$ LINER galaxies and one Seyfert II galaxy, i.e. AGNs with narrow lines.
The analysis of the Seyfert II galaxy in our sample shows that the gravitational redshift of its lines is compatible with zero, which is consistent with emission from {a more extended} narrow line region.} 

In the case of the LINER galaxies, the observation of relatively narrow lines with such a high {gravitational} redshift 
is a puzzle that will remain unsolved in this work; although the gravitational redshift is of the expected order, the FWHM values are too small compared to those given by Eq.~\ref{1}. {Indeed, although in Fig.~\ref{correlation} we {fix} the correct slope, the intercept is two orders of magnitude higher, $f_{\rm fit}\approx 400$, than that expected in Eq. \ref{1}, $f\approx1$}. 
An interesting possibility to consider here is the case where  the galaxy nucleus is affected by  outflows due to starbursts or by radiation pressure from the BLR.  The starburst contamination {would cause a decrease 
in the resulting FWHM}. The expected decrease due to starburst driven outflows, however, is not enough to explain the low widths of our lines. Apart from a possible role played by the spatial configuration of the emission clouds, there is perhaps another mechanism related to radiation pressure.
If the clouds were close to radiation pressure equilibrium this would allow the dynamics of the BLR to show lower velocities, leading to narrower lines emitted from an otherwise broad line region. As the {acceleration due to radiation pressure} scales as $R^{-2}$, its addition to the Kepler's law (Eq.~\ref{1}) can raise considerably the factor $f$ 
if the accretion disc luminosity is close enough to the Eddington limit.

{The agreement between GAMA redshifts and our estimates of SAMI central redshifts demonstrates that the $z_g$ of a SMBH is a systematic effect present in AGN single fibre redshifts. This systematic error in the galaxy redshift can potentially affect cosmological parameter estimations that use the redshifts of SNe galaxy hosts.  However, we find that only a very small fraction of SNe hosts are Seyfert I AGNs and LINERs, indicating that this particular systematic should not affect the SNe-based estimates of the matter density and Hubble parameters.}


\begin{acknowledgements}
      We are thankful to Simone Daflon, Rosa Delgado, Luis A. D\'iaz-Garc\'ia, Renato Dupke and Antonio Hernán-Caballero for useful comments.  NDP received support from a RAICES, a RAICES Federal, and PICT-2021-I-A-00700 grants of the Ministerio de Ciencia, Tecnología e Innovación, Argentina. SC is partially supported by CNPq (Brazil) with grant 311584/2020-9. JCM acknowledges support from the European Union’s Horizon Europe research and innovation programme (COSMO-LYA, grant agreement 101044612). PC is supported by a CONICET PhD fellowship (Argentina). JSA is supported by CNPq (grant no. 307683/2022-2) and Funda\c{c}\~ao de Amparo \`a Pesquisa do Estado do Rio de Janeiro - FAPERJ (grant no. 233906).
\end{acknowledgements}

\bibliographystyle{aa}
\bibliography{biblio}

\section*{Appendix}
    
\begin{figure*}
    \centering
    \vskip -0.3cm
    \includegraphics[scale=.7]{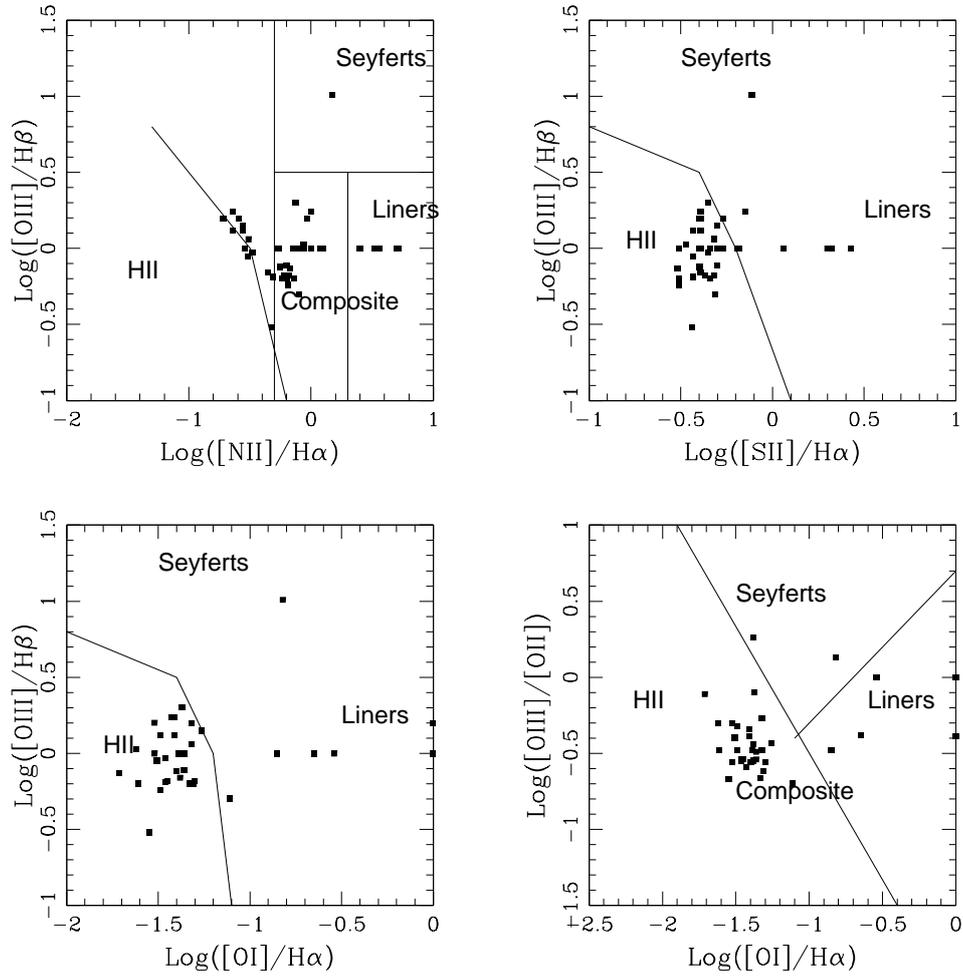}
    \vskip -3.7cm
    \caption{Classification diagrams for the SAMI galaxies used in this study according to \cite{kauffmann03}.  Cases with zero line luminosity in   $H_{\beta}$ or $[OI]$ $6300$ Angstr\"oms  are still shown, with y-axis or x-axis position set to $0$.}\label{diagram}
   \end{figure*}

Here we present the list of SAMI IDs for the galaxies used in this work. { Non AGN galaxies} (including composite and starbursts):
30916, 31452, 40283, 47460, 77452, 85474, 99349, 106616, 136263, 138047, 144239, 204799, 216670, 229167, 238125, 239376, 239560, 319020, 319400, 372374, 396607, 517247, 519089, 559061, 561020, 567624, 570090, 593680, 619737, 620034, 623432.
{ LINER galaxies:} 54198, 65406, 106549, 145729, 373202, 396621, 550328, 583427.  The Seyfert II galaxy in the sample corresponds to ID 209698. In Fig.~\ref{diagram} we show the classification diagrams.

%
%

\end{document}